\documentclass[a4paper,fleqn,usenatbib]{mnras}

\usepackage{newtxtext,newtxmath}

\usepackage[T1]{fontenc}
\usepackage{ae,aecompl}

\usepackage{graphicx}	
\usepackage{amsmath}	
\usepackage{amssymb}	

\title[Double Gravitational Wave Mergers]{Double Gravitational Wave Mergers}

\author[Samsing \& Ilan]{
Johan Samsing$^{1,2}$\thanks{E-mail: jsamsing@gmail.com},
Teva Ilan$^{1}$
\\
$^{1}$Department of Astrophysical Sciences, Princeton University, Peyton Hall, 4 Ivy Lane, Princeton, NJ 08544, USA\\
$^{2}$Einstein Fellow\\
}

\date{Accepted XXX. Received YYY; in original form ZZZ}

\pubyear{2017}

\begin{document}
\label{firstpage}
\pagerange{\pageref{firstpage}--\pageref{lastpage}}
\maketitle

\begin{abstract}
In this paper we study the dynamical outcome in which black hole (BH) binary-single interactions
lead to two successive gravitational wave (GW) mergers; a scenario we refer to as a `double GW merger'. The first GW
merger happens during the three-body interaction through a two-body GW capture, where the second GW merger
is between the BH formed in the first GW merger and the remaining bound single BH.
We estimate the probability for observing both GW mergers, and for observing only
the second merger that we refer to as a `prompt second-generation (2G) merger'.
We find that the probability for observing both GW mergers is only notable for
co-planar interactions with low GW kicks ($\lesssim 10^{1}-10^{2}$ kms$^{-1}$), which suggests that double GW mergers can be used to
probe environments facilitating such interactions. For isotropic encounters, such as the one found in globular clusters,
the probability for prompt 2G mergers to form is only at the percent level, suggesting that 
second-generation mergers are most likely to be between BHs which have swapped partners at least once.
\end{abstract}




\section{Introduction}

The recent gravitational wave (GW) detections by the Laser Interferometer Gravitational-Wave Observatory (LIGO) of
binary black hole (BBH) mergers \citep{2016PhRvL.116f1102A,
2016PhRvL.116x1103A, 2016PhRvX...6d1015A, 2017PhRvL.118v1101A, 2017PhRvL.119n1101A}, have initiated a wide range
of studies on how such BBHs form and merge in our Universe \cite[see][and references therein]{2016ApJ...818L..22A}.
The merger channels that have been suggested to contribute include
stellar clusters \citep{2000ApJ...528L..17P, 2010MNRAS.402..371B, 2013MNRAS.435.1358T, 2014MNRAS.440.2714B,
2015PhRvL.115e1101R, 2016PhRvD..93h4029R, 2016ApJ...824L...8R, 2016ApJ...824L...8R, 2017MNRAS.464L..36A, 2017MNRAS.469.4665P},
primordial black holes \citep{2016PhRvL.116t1301B, 2016PhRvD..94h4013C, 2016PhRvL.117f1101S, 2016PhRvD..94h3504C},
active galactic nuclei discs \citep{2017ApJ...835..165B,  2017MNRAS.464..946S, 2017arXiv170207818M},
galactic nuclei \citep{2009MNRAS.395.2127O, 2015MNRAS.448..754H, 2016ApJ...828...77V, 2016ApJ...831..187A, 2017arXiv170609896H},
isolated field binaries \citep{2012ApJ...759...52D, 2013ApJ...779...72D, 2015ApJ...806..263D, 2016ApJ...819..108B, 2016Natur.534..512B},
and field triples \citep[e.g.][]{2017ApJ...836...39S, 2018arXiv180508212R}.

Theoretical work indicates that these channels likely result in similar merger rates and chirp mass distributions \citep[e.g.][]{2017ApJ...846...82Z},
which has lead to several discussions on how to observationally distinguish them. From this, it seems that the spins of the BHs and their orbital
eccentricity in the LIGO band, are very promising parameters. Regarding spins, BH spins
are naturally believed to be isotropically distributed for the dynamical channel \citep[e.g.][]{2016ApJ...832L...2R, 2017arXiv170601385F}, whereas the field binary channel
is more likely to result in BH spins that are preferentially aligned \citep{2000ApJ...541..319K}. In case of eccentricity, BBH mergers from the field binary channel
are expected to be near circular at the time of observation \citep[e.g.][]{2003MNRAS.342.1169V}, whereas a notable fraction of the dynamically assembled
BBH mergers are likely to be eccentric in the LIGO band as a result of chaotic exchanges of angular momentum
during their formation \citep[e.g.][]{2006ApJ...640..156G, 2014ApJ...784...71S, 2016ApJ...816...65A,
2017ApJ...846...36S, 2017ApJ...840L..14S, 2017arXiv171107452S, 2018ApJ...853..140S}.

In this paper we study the role of post-Newtonian (PN) effects \citep[e.g.][]{Blanchet:2006kp} -- or GR corrections -- in few-body interactions,
for exploring the interesting and distinct outcome in which a binary-single interaction results in not only one,
but in {\it two} GW mergers; an outcome first studied numerically by \cite{2008PhRvD..77j1501C}, and later dynamically by \cite{2018MNRAS.476.1548S}.
To shorten the descriptions, we denote this outcome a \emph{double GW merger}. In this scenario, the first
GW merger forms through a standard two-body GW capture \citep[e.g.][]{Hansen:1972il, 2017arXiv170101548B}, while
the three BHs temporary constitute a bound state \citep[e.g.][]{2014ApJ...784...71S}. After this follows the formation
of the second GW merger, which is between the BH formed in the first GW merger and the remaining
bound BH. The endstate of the double GW merger scenario is therefore a single BH with a mass approximately that of the three initial BHs,
and a velocity composed of the initial three-body center-of-mass (COM) velocity and any acquired GW kick velocity.
We note here that this multi GW merger scenario is somewhat the GR equivalent of the classical multi stellar
collision scenario described in, e.g., \cite{Fregeau:2004fj}.

In the work presented here, we explore two unique observables related to the double GW merger scenario:
i) The first is the observation of both GW mergers, which naturally requires that the time between the
first and the second GW merger, denoted $t_{12}$, is less than the observation time window ($< \mathcal{O}(10^{1})$ years). 
For this outcome, we find that near co-planar interactions
generally give rise to the shortest time interval $t_{12}$, which in a few cases is $ < \mathcal{O}{(\text{years})}$.
The reason for a co-planar preference is that in this case the angular momentum carried by the incoming single can significantly reduce
the angular momentum of the initial target binary and thereby the GW life time.
This makes the double GW merger channel an indirect probe of environments facilitating co-planar interactions.
Although speculative, such environments may include rotationally supported systems, such as an
active galactic nuclei (AGN) disk \citep[e.g.][]{2016ApJ...819L..17B, 2017MNRAS.464..946S, 2017ApJ...835..165B, 2017arXiv170207818M,
2018MNRAS.474.5672L}. Another example is the class of disk-like systems forming in galactic nuclei as a result of vector resonant
relaxation \cite[e.g.][]{2011MNRAS.412..187K, 2015MNRAS.448.3265K, 2017ApJ...842...90R}.
ii) The second is the observation of only the second GW merger, which we in short will refer to as prompt second-generation (2G) merger.
As most binary-single interactions are expected to be between objects of similar mass \citep[][]{2016PhRvD..93h4029R, 2017ApJ...840L..14S}, prompt
2G mergers will most often be between two BHs with mass ratio about $1:2$, where the heavier BH spins at $\sim 0.7$.
Recent studies have suggested that dynamical environments indirectly can be probed by the search for
second-generation mergers \citep[e.g.][]{2016ApJ...831..187A, 2017PhRvD..95l4046G, 2018PhRvL.120o1101R}; however, the contribution
from our presented prompt 2G channel has not been discussed
in the literature so far. For a 2G merger to occur the time span $t_{12}$ has to be smaller than the average encounter time scale for the
considered dynamical system ($< 10^{8}$ years). In this case we also find that near co-planar interactions can significantly contribute to a prompt 2G population,
whereas the rate of prompt 2G mergers in isotropic systems is suppressed.

The paper is structured as follows. In Section \ref{sec:Black Hole Binary-Single Interactions} we describe some of the
basic properties related to binary-single interactions with and without GR corrections. This includes a definition of the relevant binary-single
outcomes, and a description of how these distribute in orbital phase space. In Section \ref{sec:Formation of Black Hole Double GW Mergers}
we present our main results on the formation of double GW mergers. We summarize our findings in Section \ref{sec:Discussion}.

\section{Black Hole Binary-Single Interactions}\label{sec:Black Hole Binary-Single Interactions}

The initial total energy of a binary-single interaction broadly determines the range of possible outcomes
and associated observables \citep[e.g.][]{Heggie:1975uy, Hut:1983js, 2014ApJ...784...71S}.
If the total energy is positive, also known as the soft-binary (SB) limit, the interaction is always prompt, and will most likely
lead to either an {\it ionization}, a {\it fly-by}, or an {\it exchange} endstate \citep[e.g.][]{Hut:1983js, Hut:1983by}.
If the total energy is instead negative, also known as the hard-binary (HB) limit, the system can enter a bound state with a
lifetime that generally is in the range of a few to several thousand initial orbital times \citep[e.g.][]{Hut:1983js, Hut:1993gs}.
Such bound states often undergo highly chaotic evolutions under which two of the three objects have a relative
high chance of merging \citep[e.g.][]{Fregeau:2004fj, 2014ApJ...784...71S, 2017ApJ...846...36S, 2017ApJ...840L..14S}. This makes
the HB limit interesting and highly relevant for the assembly of BBH mergers observable by
LIGO \citep[e.g.][]{2006ApJ...640..156G, 2014ApJ...784...71S, 2016PhRvD..93h4029R, 2017ApJ...840L..14S}. We will in this paper therefore solely focus on
the dynamics and outcomes of the HB limit.

\begin{figure}
\centering
\includegraphics[width=\columnwidth]{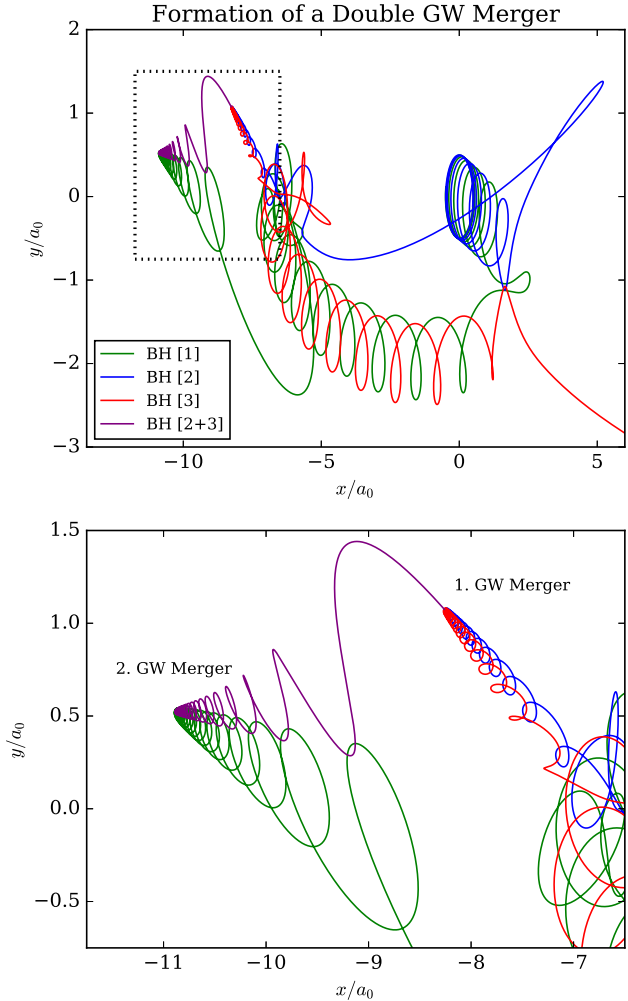}
\caption{Dynamical formation of a double GW merger.
{\it Top:} Orbital trajectories from an equal mass co-planar interaction between a BH binary and an incoming single BH, evolved using our $N$-body code that includes
GW emission in the EOM at the 2.5PN level \citep{2017ApJ...846...36S}. Each of the three BHs has a mass of $20M_{\odot}$, and the initial SMA is set to the low value of $10^{-4}$ AU
for illustrative purposes.
{\it Bottom:} Zoom in on the orbital parts showing the first GW merger (labeled `1. GW Merger'), and the second GW merger (labeled `2. GW Merger').
The zoom box is shown in the top plot by a {\it dotted box}.
As seen, the first GW merger happens here between BH[2] and BH[3] as a result of a highly eccentric GW capture \citep[e.g.][]{2014ApJ...784...71S}, from which a new BH,
denoted by BH[2+3], is formed. This first GW merger happens while the three BHs are still bound to each other,
which implies that BH[2+3] and the remaining single BH[1] are also bound to each other after the formation of BH[2+3] -- given the GW kick velocity of BH[2+3]
is negligible. In that limit, the system now undergoes its second GW merger, which is between BH[2+3] and BH[1]. For the simulation results shown here, we have for simplicity
assumed that the GW kick velocity of BH[2+3] is zero, and the mass of BH[2+3] is the mass of BH[1] plus the mass of BH[2].
As described in Section \ref{sec:g0vk0}, the time span from first to second GW merger takes it shortest values for binary-single interactions
where the total angular momentum is close to zero, a scenario that is only possible in near co-planar interactions.
An observed population of double GW mergers will therefore be an indirect probe of environments facilitating near co-planar binary-single interactions,
such as disk-like environments.}
\label{fig:DME_1}
\end{figure}

\subsection{Hard-Binary Interactions}\label{sec:Hard-Binary Limit}

A HB interaction either undergoes a direct interaction (DI) or a resonant interaction (RI), depending on the exact
ICs \citep[e.g.][]{Hut:1983js, 2014ApJ...784...71S, 2018MNRAS.476.1548S}. These two interaction types are briefly discussed below,
together with their relation to finite size effects and GR corrections.

A DI is characterized by having a relative short interaction time. The associated kinematics are generally such
that after the single enters the binary, it directly pairs up with one of the members by ejecting the remaining member to infinity
through a classical sling-shot maneuver \citep[e.g.][]{2018MNRAS.476.1548S}. Due to the short nature of this interaction type, finite sizes and GR effects rarely
play a role here \citep[e.g.][]{2014ApJ...784...71S}.

A RI is by contrast characterized by having a long interaction time, as the three objects in this case
enter a temporary bound state. The associated evolution of this state is often highly chaotic \citep[e.g.][]{Hut:1983js, 2014ApJ...784...71S}, which 
makes it possible for the system to enter a configuration where dissipative effects, such as GW emission and tides, become important
for the subsequent dynamics \citep{2017ApJ...846...36S}. To understand how, it was illustrated
in, e.g., \cite{2014ApJ...784...71S, 2017ApJ...846...36S, 2018ApJ...853..140S} that a typical RI
can be described as a series of intermediate states (IMSs) characterized by a binary, referred to as the IMS binary, with
a bound single. Between each IMS the three objects undergo a strong interaction where they semi-randomly exchange energy and angular momentum.
Each IMS binary is therefore formed with orbital parameters that are different from that of the initial target binary, which
makes it possible for highly eccentric IMS binaries to form during the interaction. It is primarily during the evolution of these highly eccentric
IMS binaries that GW emission can drain a notable amount of orbital energy and angular momentum out of the three-body system.
The change in outcome distributions from including GW emission in the EOM, is therefore linked to the
formation of highly eccentric IMS binaries in RIs. This will be described further below.

\subsection{Outcomes and Endstates}\label{sec:Outcomes and Endstates}

In the classical case where the objects are assumed point-like and only Newtonian gravity is included in the $N$-body EOM,
the only possible endstate in the HB limit is a binary with an unbound single \citep[e.g.][]{Heggie:1975uy, Hut:1983js}. We refer to this endstate binary as the post-interaction binary
in analogy with \cite{2017ApJ...840L..14S, 2018ApJ...853..140S}. In this notation, if the initial incoming single is a member of the post-interaction binary the endstate will be an exchange.

If one includes finite sizes, a collision between any two of the three objects become a possible endstate.
Such collisions predominately form in RIs, as a result of an IMS binary forming with a pericenter distance that is smaller than the
sum of the radii of the two IMS binary members \citep[e.g.][]{Fregeau:2004fj, 2017ApJ...846...36S}. The probability for a
post-interaction binary to undergo a collision is in comparison relatively small,
as each binary-single interaction leads to several IMS binaries compared to only one post-interaction binary.
As a result, the majority of the collisional products forming in binary-single interactions will therefore have the remaining single as bound companion.
As argued in \cite{2017ApJ...846...36S}, collisions contribute significantly to the merger rate when the
interacting objects are solar type objects, and less if they are compact, such as a white-dwarf, a neutron-star (NS) or a BH. In the latter case, dissipative captures (tides and GWs)
greatly dominate over collisions \citep{2018ApJ...853..140S}.

If GW emission is included in the EOM (in our simulations through the 2.5PN term), a close passage between any two of the three objects can lead to a significant
loss of orbital energy and angular momentum \citep[e.g.][]{2006ApJ...640..156G}. If the energy loss is large enough, the two objects will undergo a GW capture
with a merger to follow, which we in the three-body case refer to as a {\it GW inspiral} in analogy with \cite{2014ApJ...784...71S}.
This GW inspiral scenario most often happens between IMS binary members, as these have a finite probability for being formed with a
high eccentricity and thereby small pericenter distance, as described in Section \ref{sec:Hard-Binary Limit}.
As demonstrated by \cite{2014ApJ...784...71S}, the probability for forming a GW inspiral can
therefore be estimated from simply deriving the fraction of IMS binaries that form with a GW inspiral time that is shorter
than the orbital time of the bound single. As this GW inspiral time has to be comparable to the orbital time of the initial binary, the IMS binary eccentricity
must be close to unity, which explains why the binary-single channel is a natural producer of high eccentricity BBH mergers \citep{2017ApJ...840L..14S, 2017arXiv171107452S}.

Figure \ref{fig:DME_1} shows an example of a GW inspiral forming during a binary-single interaction.
As for the collisions, in the limit where GW velocity kicks are negligible, the merger remnant formed as a result of the GW inspiral will still be bound to
the remaining single after its formation. This final binary, now composed of the single and the merger product, will undergo its own GW merger on a
timescale that depends sensitively on the ICs, as will be described in Section \ref{sec:Formation of Black Hole Double GW Mergers}. It is the inspiral time
of this second GW merger that ultimately determines if the double GW merger scenario is observable or not.

\subsection{Distribution of Endstates}

\begin{figure}
\centering
\includegraphics[width=\columnwidth]{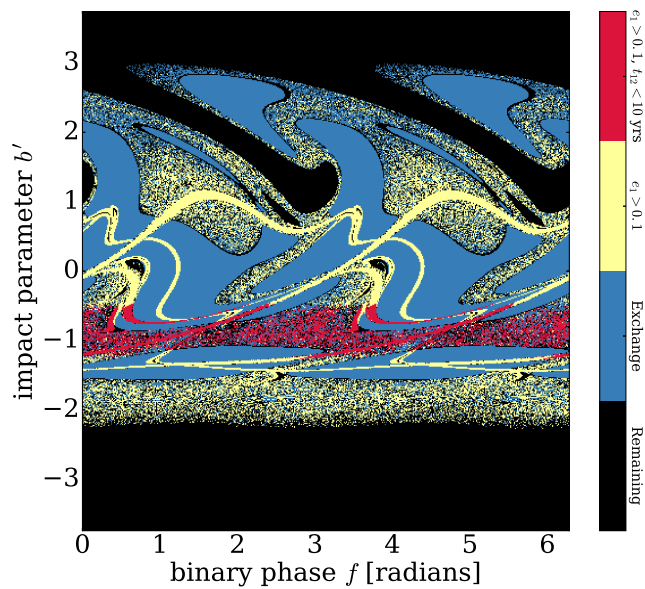}
\caption{Endstate topology.
Distribution of binary-single endstates, as a function of the binary phase $f$, measured at the time
the single is exactly $20 \times a_{0}$ away from the binary COM, and the rescaled impact
parameter $b' = (b/a_{0})(v_{\infty}/v_{\rm c})$ (see Section \ref{sec:Second GW Merger}). We refer to this endstate
distribution as the endstate topology. The shown distribution is derived from equal mass co-planar binary-single BH interactions ($\gamma = 0$), where each BH has a
mass of $20M_{\odot}$, and the SMA of the initial target binary is $a_{0} = 10^{-2}$ AU.
{\it Blue:} All interactions resulting in a classical exchange interaction.
{\it Yellow:} Interactions that result in a merging BH binary that has an orbital eccentricity $e_{1} > 0.1$ when its GW peak frequency is at $50$ Hz. This
is the population that is likely to appear eccentric when observed by an instrument similar to LIGO.
{\it Red:} Interactions from the yellow population ($e_{1}>0.1$ at $50$ Hz) that further have a
timespan from first to second GW merger $t_{12}<10$ years. We have here assumed a GW kick velocity of $v_{\rm K} = 0$.
{\it Black:} All remaining endstates and unfinished interactions.
How the fraction of the potential observable red population ($e_{1}>0.1$ at $50$ Hz with $t_{12}<10$ years)
scales with the initial SMA $a_{0}$ and the GW kick velocity $v_{\rm K}$, is shown in Figure \ref{fig:prob_DM}
and discussed in Section \ref{sec:Formation of Black Hole Double GW Mergers}.
}
\label{fig:TOP_1}
\end{figure}

To understand which binary-single ICs lead to exchanges, collisions, single and double GW mergers,
we now consider a mapping we refer to as the endstate topology \citep{1983AJ.....88.1549H, 2018MNRAS.476.1548S}. This topological mapping refers to
the graphical representation of the distribution of endstates, as a function of the initial binary phase $f$ and impact parameter $b$, as further
described in Figure \ref{fig:TOP_1} and \ref{fig:ICs}, and \cite{1983AJ.....88.1549H, 2018MNRAS.476.1548S}. We note here that for isotropic encounters
each point in the $f,b$ space is equally likely.

The endstate topology derived for a co-planar equal mass BH binary-single interaction is shown in Figure \ref{fig:TOP_1},
where {\it blue} denotes an exchange endstate, {\it yellow} all endstates for which
the endstate binary (that eventually merges due to GW emission) has an eccentricity $e_{1} > 0.1$ when its GW peak frequency $f_{\rm 1GW}$ is at $50$ Hz,
and {\it black} all remaining endstates and unfinished interactions (the {\it red}
relates to double GW mergers and will be explained in Section \ref{sec:Second GW Merger}).
For calculating the eccentricity distribution at $50$ Hz, we evolved the endstate binary in
question using the quadrupole formalism from \cite{Peters:1964bc}, together with $f_{\rm 1GW}$ using the approximation
presented in \cite{Wen:2003bu}.  We note that the vast majority of the highly eccentric GW inspiral mergers originate from IMS binaries
undergoing either a GW inspiral or a direct collision.

As seen in Figure \ref{fig:TOP_1}, the distribution of endstates is far from random, despite the
chaotic nature of the three-body problem \cite[e.g.][]{1983AJ.....88.1549H, 2018MNRAS.476.1548S}.
Generally, the large-scale wave-like pattern arises from the distribution of how much energy the single ejected during the first
sling-shot maneuver receives (see Section \ref{sec:Hard-Binary Limit}), as the homogenous regions (neighboring $f,b$ points have the same endstate) arise
from DIs and the random regions (neighboring $f,b$ points have semi-random endstates) arise from RIs \citep[e.g.][]{2018MNRAS.476.1548S}.
The asymmetry along the $b$-axis relates to which way the binary rotates relative to the incoming single, where $b>0$ corresponds to prograde motion,
and $b<0$ to retrograde motion. Also, the initial total angular momentum, $L_{\rm 0}$, changes along the $b$-axis, as the
angular momentum brought in by the single is $\propto b$. As discussed in the sections below,
understanding how $L_{\rm 0}$ distributes is the key to understand what ICs that will lead to double GW mergers
with a time space $t_{12}$ short enough for either resulting in two observable mergers, or a single prompt 2G merger.

\section{Formation of Double GW Mergers}\label{sec:Formation of Black Hole Double GW Mergers}

Having provided an understanding of binary-single interactions with
GR and finite size effects, we are now in a position to study our proposed double GW
merger scenario. We proceed below by first describing the dynamics leading to the
two GW mergers (Section \ref{sec:First GW Merger} and \ref{sec:Second GW Merger}), after which we discuss
the prospects for observing either both GW mergers (Section \ref{sec:Time from First to Second GW Merger}), or
just the second prompt 2G merger (Section \ref{sec:Formation of Prompt 2G Mergers}).
For this, we use analytical relations together with full numerical 2.5PN simulations including GW kicks.
We note here that at least one of the two GW mergers is likely to have a notable eccentricity when entering the LIGO band,
observing double GW mergers therefore heavily relies on the development of accurate eccentric GW
templates \citep[e.g.][]{2016PhRvD..94b4012H, 2017PhRvD..95b4038H, 2017arXiv170510781G, 2018PhRvD..97b4031H}.

\subsection{First GW Merger}\label{sec:First GW Merger}

The first GW merger forms through a GW inspiral, while the three objects are still bound to each other, as described in Section \ref{sec:Outcomes and Endstates}.
The resultant BH has a mass close to the total mass of the two merging BHs (ignoring relativistic mass loss), and a velocity that is composed of the initial
COM velocity of the two merging BHs, and a GW kick velocity gained through asymmetric GW emission at
merger \citep[e.g.][]{2006ApJ...653L..93B, 2007PhRvL..98i1101G, 2014ARA&A..52..661L}.
The magnitude of the GW kick velocity depends on the relative BH spins and masses,
and can easily reach values of order $\sim 10^{2} -10^{3}\ \text{km s}^{-1}$ \citep[e.g.][]{2010RvMP...82.3069C}. The GW kick velocity can therefore significantly
affect the dynamics leading to the second GW merger, and plays as a result an important role. We study the effect from GW velocity kicks
in Section \ref{sec:gammaEQ0vkGT0} (co-planar case) and Section \ref{sec:ggt0vk0} (isotropic case).
As the first GW merger generally enters the LIGO band with a notable eccentricity \citep{2014ApJ...784...71S, 2017ApJ...840L..14S}, 
an observation of an eccentric BBH merger is therefore the best indication of a potential double GW merger.

\subsection{Second GW Merger}\label{sec:Second GW Merger}

\begin{figure}
\centering
\includegraphics[width=\columnwidth]{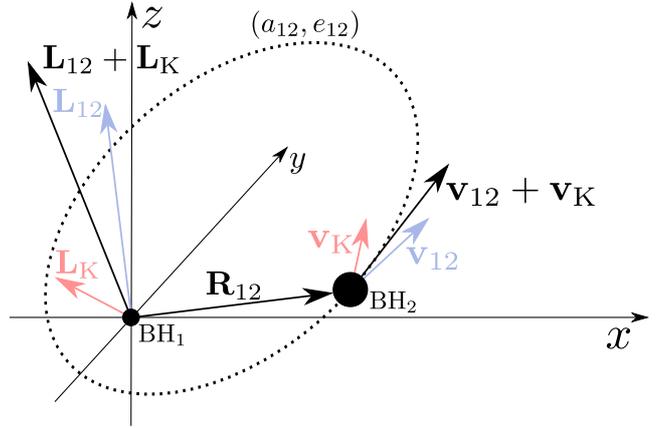}
\caption{Illustration of the orbital configuration right after the first GW merger.
As described in Section \ref{sec:First GW Merger}, the first GW merger happens through a two-body GW capture, while the three BHs
are still bound to each other. The BH formed as a result of this first GW merger is labeled above by `BH$_{2}$', and the remaining
single by `BH$_{1}$'. The relative position vector pointing from BH$_{1}$ to BH$_{2}$ is labeled by ${\bf R}_{12}$.
In the frame of BH$_{1}$, BH$_{2}$ will right after its formation move with a velocity that is composed of the COM velocity of the two BHs that merged to form BH$_{2}$, and the
GW kick velocity gained in the GW merger. The COM velocity is labeled by ${\bf v}_{12}$ ({\it light blue}), the GW kick velocity
by ${\bf v}_{\rm K}$ ({\it light red}), and the corresponding angular momentum vectors by ${\bf L}_{\rm 12}$ ({\it light blue}) and ${\bf L}_{\rm K}$ ({\it light red}), respectively.
Right after the first GW merger, BH$_{2}$ therefore moves with a velocity ${\bf v}_{12} + {\bf v}_{\rm K}$ ({\it solid black}) on an orbit with corresponding
angular momentum ${\bf L}_{\rm 12} + {\bf L}_{\rm K}$ ({\it solid black}). The SMA and eccentricity of this orbit ({\it black dotted line}) is
labeled by $a_{12}, e_{12}$, respectively. If the GW kick velocity is not unbinding the system, then BH$_{1}$ and BH$_{2}$ will undergo a GW inspiral
on a timescale $t_{12}$ that is set by the orbital parameters $a_{12}, e_{12}$. If $t_{12}$ is short, then both the first and the second GW merger can be observed.}
\label{fig:secondM_conf}
\end{figure}

The second GW merger forms through an inspiral between the remaining bound single BH and the 
BH formed in the first GW merger. We refer in the following to these two BHs as BH$_{\rm 1}$ and BH$_{\rm 2}$, respectively.
The SMA, $a_{12}$, and eccentricity, $e_{12}$, of this [BH$_{\rm 1}$, BH$_{\rm 2}$] binary measured just after the
formation of BH$_{\rm 2}$, can be written by the use of standard classical mechanics as,
\begin{equation}
a_{12} = \left[ \frac{2}{|{\bf R}_{12}|} - \frac{\left({\bf v}_{\rm 12} + {\bf v}_{\rm K} \right)^{2}}{3Gm_{\rm BH}} \right]^{-1},
\end{equation}
\begin{equation}
e_{12} = \left[ 1 - \frac{3}{4} \frac{|{\bf L}_{12} + {\bf L}_{\rm K}|^{2}}{Gm_{\rm BH}^{3}a_{12}}\right]^{1/2},
\end{equation}
where $m_{\rm BH}$ is the mass of one of the three initial (equal mass) BHs, ${\bf v}_{\rm K}$ is the GW velocity kick vector in the frame of BH$_{\rm 2}$,
${\bf R}_{12}$ and ${\bf v}_{\rm 12}$ are the position and velocity vectors of BH$_{\rm 2}$ in the frame of BH$_{\rm 1}$
assuming ${\bf v}_{\rm K} = 0$, and ${\bf L}_{12}$ and ${\bf L}_{\rm K}$ are the
angular momentum vectors $(2m_{\rm BH}/3){\bf R_{\rm 12}} \times {\bf v}_{\rm 12}$, $(2m_{\rm BH}/3){\bf R_{\rm 12}} \times {\bf v}_{\rm K}$, respectively.
We have in these expressions assumed for simplicity that the mass of BH$_{\rm 2}$ is $ = 2m_{\rm BH}$, although the actual mass will be a few percent lower due to relativistic
mass loss at merger \citep[e.g.][]{2014ARA&A..52..661L}. An illustration of the orbital configuration right after the formation of BH$_{2}$ is shown in Figure \ref{fig:secondM_conf}.

\begin{figure}
\centering
\includegraphics[width=\columnwidth]{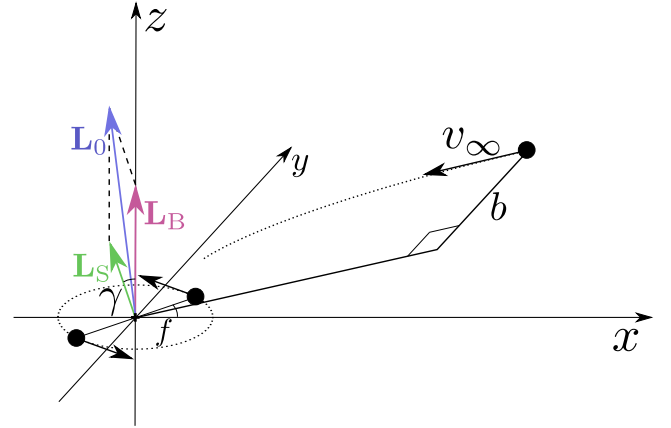}
\caption{Illustration of the binary-single ICs.
The three large {\it black dots} represent the interacting BHs, two of which initially are in a binary that here is centered at $(0,0,0)$.
The impact parameter $b$ and relative velocity $v_{\infty}$ are both defined at infinity, where $f$ refers to the binary phase at the time the single is exactly $20 \times a_{0}$ away from
the binary COM \citep{2018MNRAS.476.1548S}.
The initial angular momentum vectors of the binary and the single, respectively, are here denoted by ${\bf L}_{\rm B}$ ({\it purple})
and ${\bf L}_{\rm S}$ ({\it green}), where their sum is labeled by ${\bf L}_{\rm 0} = {\bf L}_{\rm B} + {\bf L}_{\rm S}$ ({\it blue}).
The smallest angle between ${\bf L}_{\rm B}$ and ${\bf L}_{\rm S}$ is denoted by $\gamma$, and a co-planar interaction therefore has $\gamma = 0$.
As indicated by the {\it velocity arrows}, the binary is initially set to rotate counter-clockwise,
which implies that ${\bf L}_{\rm B} \cdot {\bf L}_{\rm S} > 0$ for $b>0$, and vice versa. From this follows that if $b<0$,
then $|{\bf L}_{\rm 0}|$ can be smaller than $|{\bf L}_{\rm B}|$, which can lead to surprisingly short time intervals between the
first and the second GW merger, as explained further in Section \ref{sec:Second GW Merger}.}
\label{fig:ICs}
\end{figure}

Using the above relations for $a_{12}$ and $e_{12}$, one finds that the corresponding pericenter
distance $r_{12}  = a_{12}(1-e_{12})$, in the high eccentricity limit, can be written as,
\begin{equation}
r_{\rm 12}  \approx \frac{3}{8}\frac{|{\bf L}_{12} + {\bf L}_{\rm K}|^{2}}{Gm_{\rm BH}^{3}}.
\label{eq:r12}
\end{equation}
While ${\bf R}_{12}$, ${\bf v}_{\rm 12}$, and ${\bf v}_{\rm K}$ depend sensitively on the ICs,
we find that the initial total three-body angular momentum, denoted ${\bf L}_{0} = {\bf L}_{\rm B} + {\bf L}_{\rm S}$, where ${\bf L}_{\rm B}$ and
${\bf L}_{\rm S}$ are the angular momentum vectors of the initial binary and the incoming single, respectively (See Figure \ref{fig:ICs}),
do not change significantly during the interaction. This follows from
that the first GW inspiral generally forms on a low angular momentum orbit, meaning that only a small amount of angular momentum
is emitted in the first GW merger.
As a result, one can write ${\bf L}_{12}^{2}$ to leading order as,
\begin{equation}
{\bf L}_{\rm 12}^{2} \approx {\bf L}_{\rm 0}^{2} = \frac{Gm_{\rm BH}^{3}a_{0}}{2} \left[ 1 + \frac{4}{3}b'^{2} + \frac{4}{\sqrt{3}}b' \cos(\gamma) \right],
\label{eq:L12}
\end{equation}
where $a_{0}$ is the SMA of the initial circular target binary, $\gamma$ is the smallest angle between ${\bf L}_{\rm B}$ and ${\bf L}_{\rm S}$ (See Figure \ref{fig:ICs}),
and $b' = ({b}/{a_{0}})({v_{\infty}}/{v_{\rm c}})$, where $v_{\rm c}$ is here the binary-single characteristic velocity
and $v_{\infty}$ is the relative velocity between the binary and single at infinity \citep{Hut:1983js}. In this notation $\gamma = 0$ corresponds to a co-planar interaction, and
${v_{\infty}}/{v_{\rm c}} < 1$ is what is referred to as the HB limit.

The orbit averaged inspiral time of the [BH$_{\rm 1}$, BH$_{\rm 2}$] binary, equivalent to the time span $t_{12}$, can at quadrupole order in the high eccentricity
limit be written as \citep{Peters:1964bc},
\begin{equation}
t_{12} \approx  \frac{c^{5}}{G^{3}} \frac{4\sqrt{2}}{85} \frac{r_{\rm 12}^{7/2}a_{12}^{1/2}}{m_{\rm BH}^{3}}.
\label{eq:t12_general}
\end{equation}
In the limit where the approximations employed in Equation \eqref{eq:r12} and \eqref{eq:L12} are valid, the time span $t_{12}$ from Equation \eqref{eq:t12_general}
can now then be expressed as,
\begin{equation}
t_{12} \approx \frac{\xi a_{0}^{4}}{m_{\rm BH}^{3}} \left(\frac{a_{12}}{a_{0}}\right)^{1/2} \frac{|{\bf L}_{0} + {\bf L}_{\rm K}|^{7}}{|{\bf L}_{\rm B}|^{7}},
\label{eq:t12_Lapp}
\end{equation}
where we have introduced the following constant,
\begin{equation}
\xi = \frac{c^{5}}{G^{3}} \frac{4\sqrt{2}}{85} \left(\frac{3}{16}\right)^{7/2}.
\end{equation}
For comparison, we note that the inspiral time of the initial circular binary, denoted $t_{0}$, can be written as \citep{Peters:1964bc},
\begin{equation}
t_{0} \approx \frac{\xi a_{0}^{4}}{m_{\rm BH}^{3}} \frac{1700\sqrt{6}}{81} \approx 2 \cdot 10^{5} \ \text{yrs} \left(\frac{a_{0}}{10^{-2}\text{AU}}\right)^{4} \left(\frac{m_{\rm BH}}{20M_{\odot}}\right)^{-3}.
\end{equation}
As seen, the normalizations of $t_{12}$ and $t_{0}$ are of similar order (small variations of ${\bf L}_{\rm K}$ easily result in variations of order the factor of
difference $1700\sqrt{6}/81 \approx 51$),
which shows the importance of reducing the angular momentum $|{\bf L}_{0}|$ for bringing the time span $t_{12}$
within either the observation time ($10^{0}-10^{1}$ years), or the binary-single encounter time ($10^{7}-10^{8}$ years).

The relation shown in Equation \eqref{eq:t12_Lapp} likewise illustrates that
$t_{12}$ will not be significantly affected by GW velocity kicks if $|{\bf L}_{\rm K}| \ll |{\bf L}_{\rm B}|$,
which approximately corresponds to the requirement $v_{\rm K} \ll v_{\rm 0}$, where $v_{0}$ denotes the orbital velocity of the initial target binary given by,
\begin{equation}
v_{0}  \approx 1883\ \text{km s}^{-1} \left(\frac{a_{0}}{10^{-2}\text{AU}}\right)^{-1/2} \left(\frac{m_{\rm BH}}{20M_{\odot}}\right)^{1/2}.
\label{eq:inibin_v0}
\end{equation}
A GW kick of order say $100$ km s$^{-1}$, is therefore not expected to impact the results significantly for the employed normalizations.
However, for interactions with a wider initial binary and lower BH masses, moderate GW kicks can easily unbind the system after which
a double GW merger is not forming.

Finally, we do note that binary-single systems with a relative high initial angular momentum ($|{\bf L}_{\rm 0}| \gtrsim |{\bf L}_{\rm B}|$)
can, under rare circumstances, still undergo a prompt double GW merger, provided the GW kick velocity vector is fine-tuned in such a way that $|{\bf L}_{0} + {\bf L}_{\rm K}| \ll |{\bf L}_{\rm B}|$.
Such situations show up in the endstate topology map shown in Figure \ref{fig:TOP_1}, by a `blurring' of the horizontal edges between the red and the yellow points. 
In Section \ref{sec:gammaEQ0vkGT0} we explore the effect from GW kicks using full numerical simulations.

\subsection{Observation of the First and Second GW Merger}\label{sec:Time from First to Second GW Merger}

We now explore the possibilities for observing both the first and the second GW merger
in our proposed double GW merger scenario. For this, we systematically study below how
the time span $t_{12}$ depends on the binary-single ICs, and the
GW velocity kick from the first merger. The ICs for observing only the second GW merger, i.e. the 2G merger, will
be discussed separately in Section \ref{sec:Formation of Prompt 2G Mergers}.
Our considered cases are described in the paragraphs below.

\subsubsection{Interactions with $\gamma = 0$, ${v}_{\rm K} = 0$}\label{sec:g0vk0}

We start by considering the scenario for which the binary-single interaction
is co-planar ($\gamma = 0$), and the GW velocity kick is negligible (${v}_{\rm K} = 0$). This represents an
idealized scenario; however, regarding the assumption of $\gamma = 0$, it has been argued that
BBHs in disk environments will have their relative orbital inclinations reduced leading to preferentially co-planar
interactions \citep{2018MNRAS.474.5672L}. Properties of this limit is described in the following.

First, the time span $t_{12}$ can be written by the use of Equation \eqref{eq:L12} and \eqref{eq:t12_Lapp} as,
\begin{equation}
t_{12} \approx \frac{\xi a_{0}^{4}}{m_{\rm BH}^{3}} \left(\frac{a_{12}}{a_{0}}\right)^{1/2} \left[1 + \frac{2}{\sqrt{3}} b' \right]^{7}.
\label{eq:t12_g0vk0}
\end{equation}
From this we see that if the incoming single encounters the initial binary with an impact parameter $b' = b'_{0}$, where $b'_{0} = - \sqrt{3}/2 \approx - 0.87$,
then the time span $t_{12} \approx 0$, implying the second GW merger happens when BH$_{\rm 1}$ and BH$_{\rm 2}$ pass through their
first pericenter passage. In this case, the second GW merger will be that of a near head-on BH collision \citep[e.g.][]{2016PhRvD..94j4020H}.
We note here that $b'_{0}$ is simply the impact parameter for which the angular momentum of the incoming single
exactly cancels the angular momentum of the initial binary \citep{2018MNRAS.476.1548S}.

In general, double GW mergers can form from interactions with impact parameters in the
range $-2 \lesssim b' \lesssim 3$, as seen by the distribution of yellow points in Figure \ref{fig:TOP_1}.
This correspondingly implies that the distribution of $t_{12}$ generally varies over several orders of magnitude, e.g.,
from Equation \eqref{eq:t12_g0vk0} we see that $t_{12}(b'=1)/t_{12}(b'=-1) \approx 10^{8}$.
To gain insight into for which values of $b'$ the second GW merger happens within a timespan of observable interest, we now rewrite
Equation \eqref{eq:t12_g0vk0} as,
\begin{equation}
\Delta{b'} \approx 0.38 \left(\frac{a_{0}}{10^{-2}\text{AU}}\right)^{-4/7} \left(\frac{m_{\rm BH}}{20M_{\odot}}\right)^{3/7} \left(\frac{t_{12}}{10\ \text{yrs}}\right)^{1/7},
\label{eq:Deltab}
\end{equation}
where we have defined $\Delta{b'} = |b'-b'_{0}|$, and omitted the factor with $(a_{12}/a_{0})$, as its power of $-1/14$ makes it unimportant
for these estimates.
For the employed normalizations, one reads from Equation \eqref{eq:Deltab} that all interactions
with $b' = b'_{0} \pm 0.38$ that result in a double GW merger will have a time span from first to second GW merger $t_{12} < 10$ years. This is
interesting as this is not a negligible part of the available phase space; as seen on Figure \ref{fig:TOP_1}, the range $b'_{0} \pm 0.38$ covers about
$30\%$ of the relevant $f,b'$ phase space for retrograde interactions ($b'<0$).
We note here that both the position and the width of the band of red points in Figure \ref{fig:TOP_1} are accurately given by $-0.87\pm 0.38$.
This excellent agreement between our analytical derivations and our 2.5PN numerical $N$-body simulations, strongly validates
our approach and results so far.

\begin{figure}
\centering
\includegraphics[width=\columnwidth]{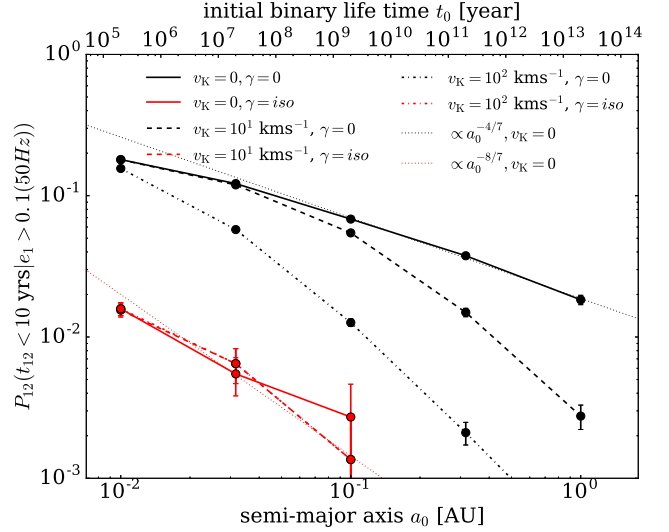}
\caption{Probability, $P_{12}$, that a BBH merger with eccentricity $e_{1}>0.1$ at $50$ Hz (first GW merger) is followed by a GW merger (second GW merger)
within a time span of $t_{12}<10$ years through our proposed double GW merger scenario. The mass of the BHs are
$m_{\rm BH} = 20 M_{\odot}$, and the probability is shown as a function of the SMA of the initial target binary, $a_{0}$.
The \emph{solid}, \emph{dashed}, and \emph{dash-dotted} lines show $P_{\rm 12}$ derived from full numerical binary-single simulations,
for when the BH formed in the first GW merger, BH$_{2}$, receives a GW velocity kick of $0$ km s$^{-1}$, $10$ km s$^{-1}$, and $100$ km s$^{-1}$, respectively.
For initial spinning BHs $v_{\rm k}$ can easily reach values of $1000$ km s$^{-1}$, which will lead to prompt disruption and no second GW merger.
For these simulations, we have assumed that the pointing of the GW kick velocity vector $\bf v_{\rm K}$ is isotropic, and the mass of
BH$_{2}$ is $2m_{\rm BH}$. The {\it black lines} show results from co-planar interactions ($\gamma = 0$), where the {\it red lines} show results from isotropic interactions ($\gamma = iso$). 
The \emph{dotted lines} show analytical scaling relations valid in the $v_{\rm K} = 0$ limit.
The \emph{top axis} shows the GW inspiral life time of the initial circular binary in years.
As seen, it is highly unlikely that isotropic environments, such as classical globular clusters, will results in observable double GW mergers.
However, co-planar interactions do generally lead to much shorter merger time scales, which correspondingly lead
to a measurable fraction of double GW mergers.
}
\label{fig:prob_DM}
\end{figure}

We now turn to the question of what the prospects are for observing both GW mergers.
For this, we performed a set of numerical binary-single interactions using our 2.5PN $N$-body code for $\gamma = 0$, $v_{\rm K} = 0$, and $m_{\rm BH} = 20M_{\odot}$,
assuming the sampling of $b'$ to be isotropic at infinity.
Results are presented in Figure \ref{fig:prob_DM}, where the black lines show the probability for that a GW merger with eccentricity $e_{1}>0.1$ at $50$ Hz (first GW merger)
will be followed by a second GW merger within a time span $t_{12}<10$ years (second GW merger) through the double GW merger scenario, as a function of the
initial binary SMA $a_{0}$.
This probability we denote by $P_{\rm 12}$ to shorten the notations. We note here that
prograde interactions ($b'>0$) will never result in short double GW mergers as this
implies that $|{\bf L}_{\rm 0}| > |{\bf L}_{\rm B}|$ and thereby $t_{12} \gtrsim t_{0}$.
From the results shown in Figure \ref{fig:prob_DM}, one concludes that for the ICs considered in this section, the probability
for that a GW merger with notable eccentricity in the LIGO band ($e_{1}>0.1$ at $50$ Hz) to have a second GW merger within
$10$ years, is $\gtrsim 0.1$ for $a_{\rm 0} < 0.1$ AU ($t_{0} \lesssim 10^{10}$ years). We note here that $\sim 0.1$ AU represents
roughly the minimum value for $a_{\rm 0}$ one would find in a classical globular cluster system; below this value the BBH would either have merged before the
next encounter, or more likely been dynamically ejected.

To gain insight into how the numerically estimated probability $P_{\rm 12}$ shown in Figure \ref{fig:prob_DM} changes with the ICs, we can use our analytical expression
for $\Delta{b'}$ from Equation \eqref{eq:Deltab}. For this, we assume that the $f,b'$ combinations for which $e_{1}>0.1$ at $50$ Hz leading to a second GW merger
are uniformly distributed near $b'_{0}$, and do not change their large scale topology when the ICs are varied. The latter assumption generally holds,
as the large scale topology results from the Newtonian `scale-free' part of the dynamics \citep{2018MNRAS.476.1548S}.
Following this approximation, the probability $P_{\rm 12}$ is simply proportional to ${\Delta}b'$.
Using our $N$-body simulations to calibrate the normalization, one now finds that the probability for 
a GW merger with eccentricity $e_{1}>0.1$ at $50$ Hz to be followed by a second GW merger within time span $t_{12}<\tau$ years, has the following
analytical solution,
\begin{equation}
P_{\rm 12} \approx 0.26 \left(\frac{a_{0}}{10^{-2}\text{AU}}\right)^{-4/7} \left(\frac{m_{\rm BH}}{20M_{\odot}}\right)^{3/7} \left(\frac{\tau}{10\ \text{yrs}}\right)^{1/7},
\label{eq:PDM}
\end{equation}
where the factor $(a_{12}/a_{0})$ again has been omitted.
Although the assumption of a uniform distribution breaks down in a few regions near $b'_{0}$ (See Figure \ref{fig:TOP_1}), we do find that
our analytical solution from the above Equation \eqref{eq:PDM} fits our numerical results quite well. As an illustrative example,
the scaling with $a_{0}$ is shown in Figure \ref{fig:prob_DM} by the {\it black dotted line}. From comparing with Figure \ref{fig:TOP_1},
the deviation from a pure powerlaw at low $a_{0}$ is indeed due to that the assumption of a uniform
distribution breaks down when $\Delta{b'}$ becomes to large, or equivalently when $a_{0}$ becomes to small.
We further note that $P_{\rm 12}$ in Equation \eqref{eq:PDM} depends very weakly on $\tau$, which implies that the
results shown in Figure \ref{fig:prob_DM} are not sensitive to the exact value of the time span limit $\tau$.
Below we study the role of GW kicks in this co-planar scenario.

\subsubsection{Interactions with $\gamma = 0$, ${v}_{\rm K} > 0$}\label{sec:gammaEQ0vkGT0}

Binary-single interactions with $\gamma \approx 0$ and $v_{\rm K} = 0$ represent the optimal scenario for producing double GW mergers with
short enough time span $t_{12}$ to be observed. However, GW kicks are expected, so here we explore how the inclusion of GW velocity kicks affects this case, by focusing on how the
fraction of double GW mergers with $t_{12}<10$ years changes with varying $v_{\rm K}$.
For this, we use our $N$-body code to numerically calculate the probability $P_{\rm 12}$, shown in Figure \ref{fig:prob_DM}, for when the BH formed in the
first GW merger, BH$_{2}$, receives a velocity kick $v_{\rm K}$ of either $10$ km s$^{-1}$ or $100$ km s$^{-1}$ at its formation.

Results for varying $v_{\rm K}$ are shown in Figure \ref{fig:prob_DM}.
As seen in the figure, GW kicks significantly reduces the fraction of double GW mergers with $t_{12}<10$ years, especially for binaries
with $a_{\rm 0}>0.1$ AU. Binaries with $a_{\rm 0}<0.1$ AU are less affected, as their orbital velocity is much higher; however, they are at the same
time less likely to exist, as they initially have a very short life time as indicated by the upper x-axis. Depending on the exact properties of their host stellar system,
very hard binaries are in fact likely to merge in-between encounters \citep[e.g.][]{2017arXiv171107452S}.
Although GW kicks are expected, we do note that several recent studies have in fact looked into the observational consequences of
BHs forming with near zero spin, which generally also leads to low kicks \citep[e.g.][]{2018PhRvL.120o1101R}.
This limit allows for the formation of second-generation GW mergers, which are characterized by mass ratios of about $1:2$
and relative high BH spins around $0.7$. In the section below we extend our analysis from this section to varying $\gamma$, including the isotropic case found
in globular clusters.

\subsubsection{Interactions with $\gamma > 0$, ${v}_{\rm K} \geq 0$}\label{sec:ggt0vk0}

We now study how varying the orbital inclination angle $\gamma$ and kick velocity $v_{\rm K}$ affect the time span $t_{12}$.
This is done to explore the role and formation probability of double GW mergers
in classical stellar systems, such as globular clusters.

To gain insight into how a varying $\gamma$ changes the picture described in Section \ref{sec:g0vk0} and \ref{sec:gammaEQ0vkGT0}, we first consider
what the minimum value for $t_{12}$ is as a function of $\gamma$, assuming the optimal case for which $v_{\rm K} = 0$.
By minimizing the expression for $t_{12}$ given by Equation \eqref{eq:t12_Lapp} w.r.t. $b'$ for fixed value of $\gamma$,
we find that the minimum value of $t_{12}$, denoted here by $\min(t_{12})$, can be written as,
\begin{equation}
\min(t_{12}) \approx \frac{\xi a_{0}^{4}}{m_{\rm BH}^{3}} \left(\frac{a_{12}}{a_{0}}\right)^{1/2} \sin^{7}(\gamma),
\label{eq:mint12}
\end{equation}
where the corresponding $b'(\min(t_{12}))$ equals $-\cos(\gamma){\sqrt{3}}/{2}$. We here see that the formation of prompt double
GW mergers with $\min(t_{12}) \approx 0$ is only possible in the special case for which $\gamma = 0$.
Considering the limit where $\sin(\gamma) \approx \gamma$, we can rewrite the above Equation \eqref{eq:mint12} in the following form,
\begin{equation}
\gamma \approx 20\degr \left(\frac{a_{0}}{10^{-2}\text{AU}}\right)^{-4/7} \left(\frac{m_{\rm BH}}{20M_{\odot}}\right)^{3/7} \left(\frac{\min(t_{12})}{10\ \text{yrs}}\right)^{1/7},
\end{equation}
where the factor $(a_{12}/a_{0})$ again has been omitted.
This relation tells us that for $a_{0} = 10^{-2} (10^{-1})$ AU interactions with $\gamma > 20\degr (5\degr)$ will not be able to result in a double GW merger
with a timespan $t_{12}<10$ years. Observable double GW mergers are therefore most likely to form in near co-planar interactions.

To study what the actual fraction of observable double GW mergers is in the isotropic case, and how much smaller it is compared to
the co-planar case, we performed $\sim 10^{6}$ $2.5$PN numerical scatterings assuming an isotropic binary-single encounter distribution.
Results are shown in Figure \ref{fig:prob_DM} with red lines. As seen, the fraction is greatly reduced, which originates from the fact that most of the
phase space ($\gamma > \mathcal{O}(10\degr)$) sampled in the isotropic case leads to a time span $t_{12}$ far too long for both mergers to be observed.
As a result, in terms of rates, isotropic environments, such as a globular cluster, are not likely to significantly contribute to
the formation of observable double GW mergers.

The probability $P_{12}$ also scales differently with $a_{0}, m_{\rm BH}$ and $\tau$ than found in the co-planar case.
The differences relate to the difference in eccentricity
distributions \citep[e.g.][]{2006tbp..book.....V, 2018MNRAS.474.5672L}; in
the isotropic case eccentricity tends to follow a thermal distribution $2e$, whereas in the co-planar case the distribution
can be shown to instead follow $e/\sqrt{1-e^2}$ \citep[e.g.][]{2006tbp..book.....V}. Assuming $e_{12}$ follow these distributions in the two scenarios,
one finds in the isotropic case that $P_{12} \propto a_{0}^{-8/7}m_{\rm BH}^{6/7}\tau^{2/7}$ and
in the co-planar case $P_{12} \propto a_{0}^{-4/7}m_{\rm BH}^{3/7}\tau^{1/7}$ (in agreement with Equation \ref{eq:PDM}).
That $P_{\rm 12}$ falls off steeper with $a_{0}$ in the isotropic case than the co-planar case,
also provides some insight into why the fraction of double GW mergers from isotropic scatterings is vanishing for
classical system for which $a_{\rm 0} > 0.1$ AU. The prospects of detecting prompt 2G mergers is described in
the section below.

\subsection{Observation of the Second Merger}\label{sec:Formation of Prompt 2G Mergers}

Having argued that {\it directly} observing both the first and the second GW merger
is only possible in near co-planar interactions and unlikely for isotropic environments,
we now turn to the question if our proposed double GW merger scenario instead can be {\it indirectly} observed. With indirect we here
refer to the case where only the prompt 2G merger is seen. As explained earlier, such a merger would be characterized by
a mass ratio of about $1:2$ and (at least) one BH highly spinning \citep[e.g.][]{2018PhRvL.120o1101R}.
For this scenario to take place, the GW life time of the second GW merger must
be shorter than the typical time between binary-single encounters, which depends on the number density of single BHs, $n_{\rm s}$, and the velocity
dispersion, $v_{\rm disp}$, of the dynamical environment as $(n_{\rm s}\sigma_{\rm bs}v_{\rm disp})^{-1}$,
where $\sigma_{\rm bs}$ is the binary-single interaction cross section  \citep[see e.g.][]{2017arXiv171107452S, 2018ApJ...853..140S}. In the following we
discuss the prospects for observing this second GW merger, i.e. the prompt 2G merger, in a typical globular cluster system in which the time between encounters
is $< 10^{8}$ years.

\begin{figure}
\centering
\includegraphics[width=\columnwidth]{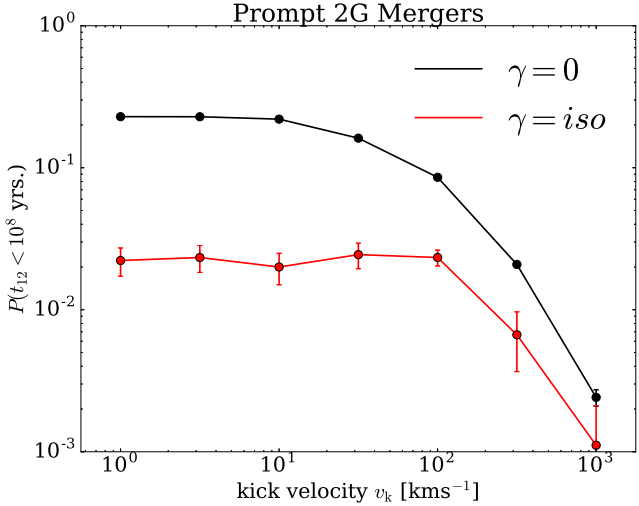}
\caption{The probability that a binary-single interaction that first undergoes a GW merger during the interaction (first GW merger)
subsequently undergoes one more GW merger (second GW merger) within a typical encounter time scale
of the host stellar system. In this figure, this time scale is set to $10^{8}$ years.
The probability is derived for a BBH with initial SMA $a_{0} = 10^{-0.5}$ AU and $m_{\rm BH} =20M_{\odot}$, as a function of the GW kick
velocity, $v_{\rm k}$. The {\it black line} shows results from co-planar interactions ($\gamma = 0$), where the {\it red line}
shows from isotropic interactions ($\gamma = iso$). As seen, only a few percent of the three-body systems that
first undergo a GW merger during the interaction will undergo a second GW merger before interacting with a new single BH.
Observations of second-generation BBH mergers forming in isotropic systems, are therefore not expected to be dominated by the
double GW merger scenario. On the other hand, a notable fraction of all co-planar interactions will undergo both the first and the second GW merger.
Such ICs also give rise to very short merger time spans $t_{12}$ as seen in Figure \ref{fig:prob_DM}. It is currently unclear if
co-planar environments exists, although they have been suggested in the recent literature \citep[e.g.][]{2018MNRAS.474.5672L}.
}
\label{fig:prob_2GM}
\end{figure}

Numerical scattering results are presented in Figure \ref{fig:prob_2GM}, which shows the probability for that the second GW merger occurs within $10^{8}$ years, for a
BBH with an initial SMA $a_{0} = 10^{-0.5} \approx 0.3$ AU and $m_{\rm BH} = 20M_{\odot}$, as a function of the GW kick velocity, $v_{\rm k}$. The black line shows the co-planar case,
where the red line shows the isotropic case. From considering the normalization of the isotropic case, we conclude that only $\sim 2\%$ of all BBH mergers
that form during three-body interactions will also undergo a second GW merger before the next encounter disrupts the system. From this we draw the following
two conclusions.

First, the [BH$_1$, BH$_2$] binary formed after the first GW merger is more likely to undergo a subsequent binary-single interaction,
than ending as a prompt 2G merger. The reason is simply that the typical time span from the first to the second GW merger, $t_{12}$, is in
the majority of cases much longer than the characteristic time between encounters.
The binary [BH$_1$, BH$_2$] formed after the first GW merger will therefore most often keep interacting and thereby contribute dynamically
to the evolution of the stellar system. We note that this is not necessarily the case for BHs formed in 2G mergers, as
the most likely outcome here is ejection as the [BH$_1$, BH$_2$] binary will be of unequal
mass with one of the BHs (highly) spinning at about $0.7$; combinations that are expected to generate large GW kicks.
A buildup and dynamical influence of third-generation BHs is therefore both difficult and unlikely.

Second, if GW mergers are observed with a notable mass ratio about $1:2$ and one BH with spin $\sim 0.7$, then the most likely origin is
a second-generation merger formed after the BH assembled in the first merger has swapped partner at least once, i.e., it is less likely that the
merger was produced in the double GW merger scenario. As described earlier, second generation BBH mergers
form either doing or in-between binary-single interactions \citep[e.g.][]{2017arXiv171107452S, 2018PhRvL.120o1101R}.
Subsequent interactions also allow for the possibility that two second-generation BHs meet each other and merge, leading to relative
heavy equal mass binaries. Such scenarios have recently been explored by e.g. \cite{2017PhRvD..95l4046G, 2018PhRvL.120o1101R},
and would point towards a dynamical origin. We conclude our study in the section below.

\section{Conclusions}\label{sec:Discussion}

We have in this paper presented a study on BH binary-single
interactions resulting in two successive GW mergers, an outcome we refer to as a
double GW merger. Double GW mergers are a natural outcome when GR effects are
included in the $N$-body EOM \citep{2018MNRAS.476.1548S}, but several mechanisms can prevent them
from happen, such as GW kicks. The formation of
double GW mergers have been proposed and presented in the past literature both numerically \citep{2008PhRvD..77j1501C}
and dynamically \citep{2018MNRAS.476.1548S}; however, our presented study is the first to quantify their actual formation
probability and in which environments they are most likely to form. Double GW mergers are interesting as they give rise to unique observables,
which could help distinguishing between BBH merger channels. A brief summery of our findings is given below.

The double GW merger scenario produces at least two unique observable signatures that are
different from other formation channels, especially the class that does not involve few-body dynamics.

The first signature, is the observation of both the first and the second GW merger in the scenario. This
requires the time span between the two mergers to be short ($t_{12} <\mathcal{O}$(years)), which is
orders of magnitude shorter than the initial target binary life time.
Using numerical and analytical arguments we have shown that only interactions that are near
co-planar will be able to produce such short double GW mergers.
The reason is that in near co-planar systems the angular momentum carried by the incoming single can
lead to a near cancelation of the initial BBH angular momentum, which correspondingly leads to a very short merger time scale \citep[e.g.][]{2018MNRAS.476.1548S}.
In the case where the encounters are instead isotropic, the overall probability for forming an observable double GW merger
decreases drastically (Figure \ref{fig:prob_DM}), as the majority of the available phase space (away from near co-planar configurations) will
lead to merger times that are far too long. From this we conclude that if both mergers in the
scenario are observed, then this would be an indirect probe of environments facilitating co-planar interactions; this includes
in particular disk systems, such as active galactic nuclei \citep[e.g.][]{2018MNRAS.474.5672L}.

The second signature, is the observation of only the second GW merger, that we refer to as a prompt 2G merger.
As pointed out in \cite{2017PhRvD..95l4046G, 2018PhRvL.120o1101R}, second-generation
BBH mergers are generally characterized by a mass ratio of about $1:2$, and at least one highly spinning BH.
Using numerical scatterings we have shown that the probability for a binary-single interaction to
undergo an observable 2G merger in the isotropic case, is still only at the percent level. Second-generation GW mergers
can form in other ways than through our proposed double GW merger scenario \citep[e.g.][]{2017PhRvD..95l4046G, 2017arXiv171107452S,
2018PhRvL.120o1101R}, and such mergers are therefore not expected to be a unique signature of the double GW merger scenario, although
they still indicate that dynamical environments are able to produce BBH mergers.

To conclude, the formation of double GW mergers have been suggested in the past literature, and their
unique observables have been numerically studied \citep[e.g.][]{2008PhRvD..77j1501C}. We have studied
their formation probability, and found that they are not expected to form in measurable numbers in classical
isotropic stellar systems, such as globular clusters. A significant fraction can only form in near co-planar environments,
which could be active galactic nuclei disks \citep[e.g.][]{2017ApJ...835..165B,  2017MNRAS.464..946S, 2017arXiv170207818M, 2018MNRAS.474.5672L}.
This makes it currently impossible to predict reliable rates, as such disk systems at present are still poorly understood.
Hopefully our study motives the community to look into this further.

\section*{Acknowledgements}
Support for this work was provided by NASA through Einstein Postdoctoral
Fellowship grant number PF4-150127 awarded by the Chandra X-ray Center, which is operated by the
Smithsonian Astrophysical Observatory for NASA under contract NAS8-03060.
TI thanks the Undergraduate Summer Research Program (USRP) at the
department of Astrophysical Sciences, Princeton University, for support.
JS thanks the Niels Bohr Institute for its hospitality
while part of this work was completed, and the Kavli Foundation and the DNRF for
supporting the 2017 Kavli Summer Program.

\bibliography{NbodyTides_papers}
\bibliographystyle{mnras}

\bsp	
\label{lastpage}
\end{document}